\documentclass[submission,copyright,creativecommons]{eptcs}
\providecommand{\event}{ICE 2015} 

\usepackage{etex}
\usepackage[usenames]{color}
\usepackage{makeidx}
\usepackage{verbatim}
\usepackage{fancyhdr}
\usepackage{fancybox}
\usepackage{stmaryrd}
\usepackage[reqno]{amsmath}
\usepackage{amsthm}
\usepackage{amssymb}
\usepackage{amsfonts}
\usepackage{amsbsy}
\usepackage{xypic}
\usepackage {indentfirst}
\usepackage{graphicx}
\usepackage[hang]{subfigure}
\usepackage{cite}
\usepackage{proof}
\usepackage{bbding}
\usepackage{pgf}
\usepackage{tikz}
\usetikzlibrary{positioning,arrows,automata}
\usepackage{cancel}
\usepackage{semantic}
\usepackage{extarrows}

 \newtheorem{theorem}{Theorem}
 \newtheorem{proposition}[theorem]{Proposition}
 \newtheorem{lemma}[theorem]{Lemma}

 \theoremstyle{definition}
 \newtheorem{definition}[theorem]{Definition}

 \theoremstyle{remark}
 \newtheorem{remark}[theorem]{Remark}

\newcommand{\HOPid}{$\Pi^d$}
\newcommand{\para}{\,|\,}
\newcommand{\fosub}[2]{\{#1/#2\} }
\newcommand{\hosub}[2]{\{#1/#2\} }
\newcommand{\ve}[1]{\widetilde{#1}}
\newcommand{\vet}[1]{{#1}}
\newcommand{\size}[1]{|#1|}
\newcommand{\st}[1]{\,{\xrightarrow{#1}}\, }

\newcommand{\wt}[1]{{\xLongrightarrow{#1}} }

\newcommand{\R}{\mathcal{R} }

\newcommand{\sepp}{\vspace*{0.3cm}}

\newcommand{\nsepv}[1]{\vspace{0mm}}
\newcommand{\SE}{\equiv }

\newcommand{\lds}{[\![}
\newcommand{\rds}{]\!]}
\newcommand{\encoding}[3]{ \lds #1 \rds^{#2}_{#3}}

\newcommand{\enc}[1]{\llbracket #1 \rrbracket}
\newcommand{\DEF}{\stackrel{\textrm{def}}{=}} 
\newcommand{\lrangle}[1]{\langle #1 \rangle} 
 
\newcommand{\PIPEA}[1]{\mathbf{#1}{\textendash}\mathfrak{pipe} }

\newcommand{\AEHOPid}{=_{\Pi^d} } 
\newcommand{\SAEHOPid}{\sim_{\Pi^d} } 
\newcommand{\AEHOPiddiv}{=_{\scriptscriptstyle \Pi^d}^{\scriptscriptstyle \Uparrow} } 
\newcommand{\AEFOPi}{=_{\pi} } 
\newcommand{\SAEFOPi}{\sim_{\pi} } 
\newcommand{\EWLB}{\approx_{l}^{\pi} } 
\newcommand{\diverge}[1]{#1^{\Uparrow}}

\makeatletter
\@ifundefined{newslide}{%
 
}{%
}
\makeatletter
\@ifundefined{endinput}{%
 
}{%
}

\newcommand{\stress}[1]{{#1}} 
\newcommand{\strs}[1]{{#1}} 

\RequirePackage[normalem]{ulem}
\RequirePackage{color}\definecolor{RED}{rgb}{1,0,0}\definecolor{BLUE}{rgb}{0,0,1}

\title{On the Computation Power of Name Parameterization in Higher-order Processes}
\author{Xian Xu
\institute{East China University of Science and Technology, China}
\email{xuxian@ecust.edu.cn}
\and
Qiang Yin \quad\qquad Huan Long
\institute{Shanghai Jiao Tong University, China}
\email{\quad yinqiang.sjtu@gmail.com \quad\qquad longhuan@sjtu.edu.cn}
}
\def\titlerunning{Name Parameterization in HO Processes}
\def\authorrunning{X. Xu, Q. Yin \& H. Long}
\begin{document}
\maketitle

\begin{abstract}
Parameterization extends higher-order processes with the capability of abstraction (akin to that in lambda-calculus), and is known to be able to enhance the expressiveness. This paper focuses on the parameterization of names, i.e. a construct that maps a name to a process, in the higher-order setting. We provide two results concerning its computation capacity. First, name parameterization brings up a complete model, in the sense that it can express an elementary interactive model with built-in recursive functions. Second,  we compare name parameterization with the well-known pi-calculus, and provide two encodings between them.
\vspace*{.1cm}

\noindent\emph{keywords}: Parameterization, Computation, Higher-order, Processes
\end{abstract}

\section{Introduction}
Parameterization stems from the \emph{abstraction} construct in the lambda-calculus by Church \cite{Bar84}. Parameterized processes are ubiquitous in modern programming languages (e.g. Java, Erlang, Clojure). A typical parameterization is the name-parameterization, 
i.e., $\lrangle{\ve{x}}P$ in which $\ve{x}$ is a sequence of parameterized names that process $P$ may take advantage of in its computation (e.g. some communication port, certain reference to database connection pool). In higher-order process models, name-parameterization has been shown to be an effective measure to increase the expressiveness, viz., passing parameterized processes (so-called abstraction-passing) is strictly more powerful than ordinary process-passing \cite{LPSS10}\cite{Xu12}\cite{XYL13}. In this paper, we deepen the study of name-parameterization in higher-order paradigm, from two perspectives: computational completeness and comparison with name-passing.

\subsubsection*{Related work and motivation}
In the field of higher-order processes, there have not been many works on computational completeness (i.e. Turing completeness) as those for first-order processes. A relatively recent notable work is by Lanese et al. \cite{LPSS10a}, who show that HOcore, a very basic higher-order process calculus with the operators of input, output and composition, is Turing complete. This significant result reveals that, in contrast to the first-order case, even with the most basic operators and without the restriction operator, higher-order processes have the computational capacity of Turing machines. Technically, the result is established by encoding Minsky machines which are equivalent to Turing machines in computation power. On the other hand, from the viewpoint of interaction, the encoding is not so strong in two aspects. Firstly, it is not compositional. That is, two processes $P,P'$ respectively encoding two Minsky machines $M,M'$ cannot interact by, for example, outputting the computing result of $P$ to $P'$ or vice versa. Secondly, the encoding is vulnerable to turbulence from the environment. That is, there are a few free (global) names visible to the observer that may easily interfere with the computation procedure. This may be largely attributed to the absence of restriction operator as well as the encoding strategy. To improve, viz. to have a robust and compositional interpretation of Turing computability, one needs more requirements on both the encoding strategy and the formulation of Turing computability (in a concurrency setting). In the first part of this paper, we try to provide one such kind of encoding.  Specifically, we show that name-parameterization in a higher-order setting grants us a compositional (and turbulence-resistent) encoding of Turing computability which is sound and complete. This kind of encoding is not likely to be achievable by HOcore. Technically, we translate an elementary interactive model called $\mathbb{C}$ \cite{Fu11a}, which encapsulates recursive functions (the computable functions) , into higher-order pi-calculus with name-parameterization (notation $\Pi^d$). In addition to providing Turing computability in a compositional way (thus also an approach of its design pattern), the encoding also brings out some basic insight into the programming capability/style of name-parameterization in the higher-order paradigm.

The second part of this paper continues to study another aspect of the computation power of name-parameterization, that is, comparison with the well-established (first-order) name-passing calculus, specifically the pi-calculus (notation $\pi$) \cite{MPW92}. Since its introduction, higher-order processes have been examined through comparison with name-passing. There are basically two directions: translation from higher-order processes to the first-order processes and the converse.  In the former direction, Sangiorgi proposes in \cite{San92} a translation from a very rich calculus, which includes both higher-order and first-order operators, and both name-parameterization and process-parameterization (i.e. abstraction on processes themselves), into $\pi$ extended also with the name-parameterization.  This translation tackles only unary parameterization, and is significant in, among others, bringing up a useful technique of triggers. Here we manage to isolate and generalize the essence of the encoding by Sangiorgi. We provide an encoding from $\Pi^d$ (which is purely higher-order) into $\pi$ (which does not have name-parameterization); moreover name-parameterization is not necessarily unary. In the latter direction, Thomsen gives in \cite{Tho93} a translation from $\pi$ to higher-order processes with the relabelling operator. This translation is important in presenting a novel way to play the role of a $\pi$ name, which is a particularly remarkable point in unveiling the difference/connection between process-passing and name-passing. However, relabelling is a rather strong operator that often arouses controversy. In contrast, parameterization is a semantically more tractable operation. So here we take advantage of the idea of Thomsen to provide an encoding of $\pi$ in $\Pi^d$.

\subsubsection*{Contribution}
In summary, the contribution of this paper is twofold.
\begin{itemize}
\item We show that higher-order processes extended by name-parameterization are capable of computing all that is computable, by providing a compositional and robust translation from $\mathbb{C}$ that is a basic concurrency model of computable functions. We show that the translation is sound and complete.
\item We compare higher-order processes extended by name-parameterization with name-passing processes, by providing two translations between them. We discuss the properties of the translations with respect to some well-known notion of encoding in the field. We show that most of the requirement are satisfied, except for the soundness of the translation from name-passing to higher-order processes with name-parameterization, in which case a weaker form of soundness is provided.
\end{itemize}
\sepp

The rest of the paper is organized as follows. Section \ref{s:definition} defines the relevant calculi. Section \ref{s:completeness} shows the result on computation completeness. Section \ref{s:expressiveness} presents the expressiveness result. Section \ref{s:conclusion} concludes the paper.


\section{Preliminary}\label{s:definition}

\subsection{Calculus $\pi$}
The first-order pi-calculus (notation $\pi$) of Milner, Parrow and Walker \cite{MPW92} is one of the most developed process calculi in the literature. Here we use a lightweight variant \cite{EN86a}\cite{EN00}, in which names (ranged over by $m,n,u,v,w$) are classified into two categories: name constants (ranged over by $a,b,c,d,e$) and name variables (ranged over by $x,y,z$). Below is the grammar whose constructs have their standard meaning. We use guarded replication instead of general replication without loss of expressiveness \cite{San98}\cite{FL09a}.
\[P,Q := 0 \,\Big{|}\, m(x).P \,\Big{|}\, \overline{m}n.P \,\Big{|}\,  (c)P \,\Big{|}\, P\para Q \,\Big{|}\, !m(x).P \,\Big{|}\, !\overline{m}n.P
\]
A name constant $a$ is bound (local or restricted) in $(a)P$, otherwise it is free (global). A name variable $x$ is bound in $a(x).P$ and free otherwise.
We use respectively $fn(\cdot), bn(\cdot), n(\cdot), fv(\cdot), bv(\cdot), v(\cdot)$ to denote free constants, bound constants, constants, free variables, bound variables, and variables in a set of processes. A name is fresh if it does not appear in any of the processes under consideration. A process $P$ is closed if $fv(P){=}\emptyset$. We consider closed processes by default. We have a few derived operators: $\overline{a}(d).P \DEF (d)\overline{a}d.P$, $a.P \DEF  a(x).P \;(x\notin fv(P))$, $\overline{a}.P \DEF \overline{a}(d).P\;(d\notin fn(P))$. A trailing $0$ process is usually trimmed. We use a tilde to stand for tuples. For $\ve{n}$, $\size{\ve{n}}$ stands for its length; $m\in \ve{n}$ means $m$ is its element; $m\ve{n}$ abbreviates adding $m$ to the tuple; $(c_1)(c_2)\cdots (c_k)E$ is written as $(c_1c_2\cdots c_k)E$ or simply $(\ve{c})E$. Substitutions $\fosub{y}{x}$ are ranged over by $\sigma$. Sometimes substituting a constant for another is called renaming, and assignment if for a variable. A context $C$ is a process with some subprocess replaced by the hole $[\cdot]$; then $C[A]$ is the process resulting from filling in the hole by process $A$.
The LTS (Labelled Transition System) comprises the rules as below (symmetric rules are skipped).
\[
\begin{array}{lll}
\infer{a(x).P\st{a(b)} P\fosub{b}{x}}{} \quad &
\infer{\overline{a}b.P\st{\overline{a}b} P}{} \quad &
\infer[{\scriptstyle bn(\lambda)\cap fn(Q)=\emptyset}]{P\para Q\st{\lambda} P'\para Q}{P\st{\lambda} P'}
\end{array}
\]
\[
\begin{array}{lll}
\infer{P\para Q\st{\tau}P'\para Q'}{P\st{a(b)}P', Q\st{\overline{a}b} Q'}\quad &
\infer{P\para Q\st{\tau}(b)(P'\para Q')}{P\st{a(b)} P', Q\st{\overline{a}(b)} Q'} \quad&
\infer[{\scriptstyle c\not\in n(\lambda)}]{(c)P\st{\lambda} P'}{P\st{\lambda} P'}
\end{array}
\]
\[
\begin{array}{lll}
\infer[{\scriptstyle c\neq a}]{(c)P\st{\overline{a}(c)} P'}{P\st{\overline{a}c} P'} \quad &
\infer{!a(x).P \st{a(b)} P\fosub{b}{x}\para !a(x).P}{}\quad  &
\infer{!\overline{a}b.P \st{\overline{a}b} P\para !\overline{a}b.P}{}
\end{array}
\]
Actions (ranged over by $\lambda$) include $\tau$, and visible ones: input ($a(b)$), output ($\overline{a}b$) and bound output ($\overline{a}(c)$) which occur on constant names, and the outputted name is a constant (i.e. no variable can be transmitted).
As usual, $\equiv$ denotes the standard structural congruence \cite{MPW92}\cite{SW01a}, which is the smallest relation satisfying the monoid laws for parallel composition, commutative laws for both composition and restriction, and a distributive law $(c)(P\para Q) \SE (c)P\para Q$ (if $c\notin fn(Q)$).
We write $\wt{}$ for the reflexive transitive closure of $\st{\tau}$, and $\wt{\lambda}$ for $\wt{}\st{\lambda}\wt{}$; then $\wt{\widehat{\lambda}}$ is $\wt{\lambda}$ if $\lambda$ is not $\tau$, and $\wt{}$ otherwise. A process $P$ is divergent, denoted $\diverge{P}$, if it has an infinite computation (i.e. $\tau$ sequence). \strs{A relation $\R$ is divergence-sensitive if whenever $P\,\R\, Q$ then $\diverge{Q}$ implies $\diverge{P}$}.
In the standard way, the bisimulation on $\pi$ is defined as below and is a congruence relation \cite{MPW92}\cite{EN00}. 
\begin{definition}[Bisimulation]\label{ex-w-bisi}
Weak bisimilarity $\AEFOPi$ is the largest symmetric bisimulation relation $\mathcal{R}$ on $\pi$ processes such that whenever $P \,\mathcal{R}\, Q$ and $P\st{\lambda} P'$ then $Q\wt{\widehat{\lambda}} Q'$ and $P' \,\mathcal{R}\, Q'$.
\end{definition}
We use $\SAEFOPi$ to denote the strong version of the bisimulation (i.e., the one obtained by replacing $\wt{\widehat{\lambda}}$ with  $\st{\lambda}$ in the definition). Notice that an alternative way to define weak bisimulation 
is to use $\wt{\lambda}$ instead of $\st{\lambda}$ in Definition \ref{ex-w-bisi} (see \cite{Mil89, SW01a}).

The local bisimilarity ($\EWLB$) characterizes weak bisimilarity, i.e. $\EWLB$ coincides with $\AEFOPi$ (see \cite{Fu05b} for a proof), and will be used when discussing the encodings. 
\begin{definition}\label{ext-local-bisi}
Local bisimilarity $\EWLB$ is the largest symmetric local bisimulation relation $\mathcal{R}$ on $\pi$ processes such that:
\begin{itemize}
\item if $P\st{\lambda} P'$, $\lambda$ is not bound output, then $Q\wt{\widehat{\lambda}} Q'$ and $P'\,\mathcal{R}\, Q'$;
\item if $P\st{\overline{a}(b)} P'$, then $Q\wt{\overline{a}(b)} Q'$, and for every $R$, $(b)(P'\para R)\,\mathcal{R}\, (b)(Q'\para R)$.
\end{itemize}
\end{definition}

\subsection{Calculus $\Pi^d$}
The grammar of higher-order pi-calculus with parameterization of names (notation \HOPid) is as below. The operators have their standard meaning \cite{San92}.
\[E,E' := 0 \;\Big{|}\; X \;\Big{|}\; a(X).E \;\Big{|}\; \overline{a}E'.E \;\Big{|}\; E\para E' \;\Big{|}\; (a)E   \;\Big{|}\; \lrangle{\ve{x}} E \;\Big{|}\; E\lrangle{\ve{m}}
\] Letters $A,B,E,F,G,P,Q,T$ range over process, and $X,Y,Z$ represent process variables.
Process of the form $\lrangle{\ve{x}} E$, in which $\ve{x}$ is not empty and bound, are \emph{parameterized processes}.
Process without outermost parameterization are \emph{non-parameterized processes}.
We have name variables, name constants and names defined in a similar way to that in $\pi$.

For abstraction $\lrangle{\ve{x}} E$ and application $E\lrangle{\ve{m}}$ to work properly, we assume a type system \cite{San92} to ensure type consistency. We will not present the typing however, because it is not important for the study in this paper (see \cite{SW01a} for a reference). Notations $fpv(\cdot)$, $bpv(\cdot)$, $pv(\cdot)$ respectively denote free process variables, bound process variables and process variables. Closed processes contain no free (name or process) variables, and are considered by default. Name substitution $E\fosub{y}{x}$ and process substitution $E\hosub{F}{X}$ are defined in the usual way, and can be extended to tuples $\fosub{\ve{n}}{\ve{m}}$ and $\hosub{\ve{E}}{\ve{X}}$. Notation $E[\ve{X}]$ is the process with possibly $\ve{X}$ occurring in it, and $E[\ve{A}]$ is  $E[\ve{X}]\{\ve{A}/\ve{X}\}$. Some CCS-like prefixes are defined as: $a$ for $a(X).0$, $\overline{a}$ for $\overline{a}0.0$. We also define below replication $!P$ and will use it in the encodings: $!P \DEF (c)(Q_c \para \overline{c}Q_c) \mbox{ where } Q_c \DEF c(X).(X\para P\para \overline{c}X)$.
In \HOPid, the structural congruence $\equiv$ (we reuse the notation) is defined similar to that of $\pi$, with one additional law for parameterization: $(\lrangle{\ve{x}}E)\lrangle{\ve{m}} \equiv E\fosub{\ve{m}}{\ve{x}}, \size{\ve{m}} {=} \size{\ve{x}}$.
The LTS rules are as below (symmetric rules omitted).
\[\begin{array}{lll}
\infer{a(X).F\st{a(E)} F\hosub{E}{X}}{} \;  &
\infer{\overline{a}E.F\st{\overline{a}E} F}{} \;  &
\infer[{\scriptstyle bn(\lambda)\cap fn(F)=\emptyset}]{E\para F\st{\lambda} E'\para F}{E\st{\lambda} E'}
\end{array}
\]
\[\begin{array}{ll}
\infer{E\para F \st{\tau}(\ve{c})(E'\,|\,F')}{E\st{a(E_1)} E' & F\st{(\ve{c})\overline{a}[E_1]} F'} \qquad&
\infer[{\scriptstyle d \in fn(E_1)-\{\ve{c},a\}}]{(d)E\st{(d)(\ve{c})\overline{a}[E_1]} E'}{E\st{(\ve{c})\overline{a}[E_1]} E'} \\
\infer[{\scriptstyle c\not\in n(\lambda)}]{(c)E\st{\lambda} (c)E'}{E\st{\lambda} E'} &
\infer{F\st{\lambda} F'}{F\equiv E, \; E\st{\lambda} E', \; E'\equiv F'}
\end{array}
\]
The transition rules are largely self-explanatory.
We use $\lambda$ for actions: $\tau$ (internal action), $a(E)$ (higher-order input), and $(\ve{x})\overline{a}E$ (higher-order output) which is sometimes written $(\ve{x})\overline{a}[E]$. Notations $\wt{}$, $\wt{\lambda}$, $\wt{\widehat{\lambda}}$ and $\diverge{P}$ (also the notion of divergence-sensitivity)  are defined in the same way as that for $\pi$.
The well-known bisimulation equivalence (and congruence) for higher-order processes is the context bisimulation \cite{San92}\cite{San94}.
\begin{definition}[Context bisimulation]
Context bisimilarity $\AEHOPid$ is the largest symmetric context bisimulation relation $\mathcal{R}$ on \HOPid~processes such that whenever $E\,\mathcal{R}\, F$, the following properties hold: %
\begin{itemize}
\item If $E \st{\lambda} E'$ and $\lambda$ is not output, then $F \wt{\widehat{\lambda}} F'$ and $E'\,\mathcal{R}\, F'$;
\item If $E \st{(\ve{c})\overline{a}E_1} E'$, then $F \wt{(\ve{d})\overline{a}E_2} F'$, and for every process $G[X]$ s.t. $\ve{c}\ve{d}\cap fn(G)=\emptyset$, it holds that $(\ve{c})(G[E_1]\para E') \; \mathcal{R}\;  (\ve{d})(G[E_2]\para F')$.
\end{itemize}
\strs{We use $\AEHOPiddiv$ to denote the divergence-sensitive context bisimilarity}.
Also $\SAEHOPid$ denotes the strong version of the context bisimilarity. 

\end{definition}

\subsection{Calculus $\mathbb{C}$}\label{sec-CModel}
The $\mathbb{C}$ calculus \cite{Fu11a} is an elementary process model with built-in computable functions. It is somewhat minimal in having the substratal computation ability (i.e. (Turing)-computable functions) and interaction ability (i.e. basic communication primitive).  In a process expression, we usually use $\underline{i}$ (respectively $\underline{x}$) to denote natural number $i$ (respectively variable $x$ of natural numbers), so as to avoid ambiguity. The processes of $\mathbb{C}$ are generated by the following BNF. 
\[
P ~:=~ 0 ~\Big{|}~ \Omega ~\Big{|}~ \overline{a}(\vet{\underline{i}}) ~\Big{|}~ F_a^b(f(\vet{\underline{x}}))  ~\Big{|}~ P\para P
\]
Intuitively, $\overline{a}(\vet{\underline{i}})$ is a process outputting natural numbers $\vet{i}$ on $a$, and $\Omega$ is a divergent process which can perform an infinite number of internal actions. Process $F_a^b(f(\vet{\underline{x}}))$, which is the encapsulation of computable function $f(\vet{\underline{x}})$, acts as a black box that gets input $\vet{\underline{m}}$ on channel $a$ and computes $f(\vet{\underline{m}})$. If $f(\vet{\underline{i}})$ is defined then it outputs $f(\vet{\underline{i}})$ over channel $b$, otherwise it diverges. Notice $f(x)$ is unary for the sake of simplicity; functions of arbitrary arity can be encoded into unary ones \cite{Cut80}. There are three types of actions: input ($a(\vet{\underline{i}})$), output ($\overline{a}(\vet{\underline{i}})$), internal move ($\tau$). We use $\lambda$ to ranger over actions. The LTS is as below (symmetric rules omitted). 
\[
\inference{}{\overline{a}(\underline{i})\st{\overline{a}(\underline{i})}0}  \quad
\inference{}{\Omega \st{\tau} \Omega } \quad
\inference{P \st{\overline{a}(\vet{\underline{i}})}P' \quad Q  \st{a(\vet{\underline{i}})}Q'}{P\para Q \st{\tau} P'|Q'} \quad
\inference{P\st{\lambda}P'}{P\para Q \st{\lambda}P'|Q}
\]
\[
\inference{f(\vet{\underline{m}}) = \underline{n}}{F_a^b(f(\vet{\underline{x}}))\st{a(\vet{\underline{m}})}\overline{b}(\underline{n})} \quad
\inference{f(\vet{\underline{m}}) \textrm{ is undefined} }{F_a^b(f(\vet{\underline{x}}))\st{a(\underline{m})}\Omega}
\]
The structural congruence $\equiv_{\mathbb{C}}$ is the least equivalence and congruence relation satisfying the following equalities.
\[
0\para  P \equiv_{\mathbb{C}} P,~  P\para Q \equiv_{\mathbb{C}} Q\para P,~  (P\para Q )\para R \equiv_{\mathbb{C}} P\para (Q\para R),~ \Omega \para \Omega \equiv_{\mathbb{C}} \Omega
\]
We shall use $\equiv_{\mathbb{C}}$ as the representative equivalence on $\mathbb{C}$.
\strs{Obviously it is divergence-sensitive.} 

We use $\simeq_{\mathbb{C}}$ to stand for the standard bisimulation with divergence-sensitivity. One important property holds for the $\mathbb{C}$-calculus is the following theorem (see \cite{Fu11a} for a proof).
\begin{theorem}\label{th-C}
$P\simeq_{\mathbb{C}}Q$ if and only if $P\equiv_{\mathbb{C}}Q$.
\end{theorem}

A process model is computation complete if it can encode $\mathbb{C}$. Note that in the definition of $\mathbb{C}$-calculus we use the abstract notion of  \emph{computable functions} \cite{Cut80}. Every function can be seen as a set of ordered pairs and henceforth is unique. In another words, in the definition of $\mathbb{C}$ we do not distinguish between different
realizations of a computable function.  Comparatively in order to encode $\mathbb{C}$, it suffices to encode the recursive realization (a.k.a., the \emph{recursive functions}) of computable functions. 


\subsection{A notion of encoding}\label{sec-cri} 
We here introduce a notion of encoding. A process model $\mathcal{L}$ is a triplet $(\mathcal{P}, \st{}, \approx)$, where $\mathcal{P}$ is the set of processes, $\st{}$ is the  LTS with a set $\mathcal{A}$ of actions, and $\approx$ is a behavioral equivalence.
Given $\mathcal{L}_i\DEF (\mathcal{P}_i, \st{}_i, \approx_i)$ ($i{=}1,2$), an encoding from  $\mathcal{L}_1$ to $\mathcal{L}_2$ is a function $\encoding{\cdot}{}{}: \mathcal{P}_1 \longrightarrow \mathcal{P}_2$ that satisfies some set of criteria.
The following criteria set used in this paper is from \cite{LPSS10} (a variant based on \cite{Gor08a}).
$\mathcal{L}_1 \sqsubseteq \mathcal{L}_2$ means there is an encoding from $\mathcal{L}_1$ to $\mathcal{L}_2$.
As shown in \cite{LPSS10}, $\sqsubseteq$ enjoys transitivity.
Notice that some encodings we are going to investigate in the paper do not satisfy all the criteria. For the sake of simplicity, we stick to the term ``encoding", and explicitly point out the properties that are not met (if any).
\begin{definition}[Criteria for encodings]\label{gorla-like-cond}
There are two categories.
\begin{enumerate}
\item[] \textbf{Static criteria}:\\
(1) Compositionality. \emph{For any $k$-ary operator $op$ of $\mathcal{L}_1$, and all $P_1,...,P_k\in \mathcal{P}_1$,  $\encoding{op(P_1,...,P_k)}{}{}$ $=\, C_{op}[\encoding{P_1}{}{},...,\encoding{P_k}{}{}]$ for some context $C_{op}[\cdots]\in \mathcal{P}_2$;} \\
(2) Name invariance. \emph{For any injective substitution $\sigma$ of names, $\encoding{P\sigma}{}{} \,=\, \encoding{P}{}{}\sigma$. }

\item[] \textbf{Dynamic criteria}: \\
(1) Forth operational correspondence. 
\emph{Whenever $P\wt{\lambda} P'$, it holds $\encoding{P}{}{} \wt{\lambda'} \approx_2 \encoding{P'}{}{}$, for some action $\lambda'$ with the same \emph{subject}  as that of $\lambda$ (the subject of an action (e.g., $a(A)$) is the name on which the action happens (e.g., $a$)) ;} \\
(2) Back operational correspondence. 
\emph{Whenever $\encoding{P}{}{} \wt{\lambda'} T$, there exist $P'$ and $\lambda$ with the same \emph{subject} as that of $\lambda'$ s.t. $P\wt{\lambda} P'$ and $T\wt{} \approx_2 \encoding{P'}{}{}$;}\\
(3) Adequacy. \emph{$P \approx_1 P'$ implies $\encoding{P}{}{} \approx_2 \encoding{P'}{}{}$. This is also known as \emph{soundness}. The converse is known as \emph{completeness};} \\
(4) Divergence-reflecting. \emph{If $\encoding{P}{}{}$ diverges, so does $P$.}
\end{enumerate}
\end{definition}



\section{Computation completeness} \label{s:completeness}
In this section we focus on interpreting $\mathbb{C}$ in $\Pi^d$. In order to do that we start by
interpreting the natural numbers. Then we go ahead to interpret the processes encapsulating all the computable
functions.

\vspace*{-4mm}
\subsection{Natural numbers}
A natural number in $\Pi^d$ is coded as a binary abstraction with two parameters. The first parameter is used to do subtraction while the second one is used to test whether the given number is zero. The definition of natural numbers is given as follows.
\[ \enc{\underline{0}}  \DEF  \lrangle{x, y}\overline{y}0 \quad \quad \enc{\underline{{n{+}1}}} \DEF \lrangle{x, y} \overline{x}\enc{\underline{n}} 
\]
The processes of zero, output and composition are translated as below.
\[
\enc{0} \,\DEF\, 0 \qquad
\enc{\Omega} \,\DEF\, !\tau \qquad
\enc{\overline{a}(\underline{n})} \,\DEF\, \overline{a}\enc{\underline{n}} \qquad
\enc{P\para Q} \,\DEF\, \enc{P}\para \enc{Q}
\]

\subsection{Recursive functions}\label{sec-recursive}
By Church-Turing Thesis, the set of (unary) computable functions coincides with the set
of all (unary) recursive functions. In the attempt to show that $\Pi^{d}$ is a computation complete model,
it is sufficient to provide an interpretation of the three basic functions: zero
function, successor function and projection function, and  three operators: composition,
recursion and minimization. We then prove that the interpretation satisfies all the requirements in Definition \ref{gorla-like-cond}, except the back operational correspondence.

\begin{enumerate}
\item Zero function: $Z_{a}^{b} \DEF a(X).\overline{b}\enc{\underline{0}}$.

\item Successor function: $Suc_{a}^{b} \DEF a(X).\overline{b}[{\lrangle{x,y}\overline{x} X}]$.

\item Projection function: $Pr(i)_{a_1,\dots, a_n}^{b} \DEF a_1(X_1).a_2(X_2)\dots a_n(X_n).\overline{b}X_i$.

\item Composition $Cmp(F,\ve{G})_{a_1,\dots, a_n}^b$: Let $F(x_1,\ldots, x_k)$ be a $k$-ary recursive function, and $^i\!G$ $(1\leq i\leq k)$ be $n$-ary recursive functions. The interpretation of $F(^1\!G,^2\!G,...,^k\!G)$, i.e. $Cmp(F,\ve{G})_{a_1,\dots, a_n}^b$ is as follows:
\[
\begin{array}{cl}\hline
& \qquad  Cmp(F,\ve{G})_{a_1\dots, a_n}^{b} \DEF \\
& \qquad (c_{11} \cdots c_{1n},\dots c_{k1} \dots c_{kn}, b_1\dots b_k) \\
\mbox{\small { \emph{initialize}}} & \qquad a_1(X_1).a_2(X_2).\cdots.a_n(X_{n}).(\\
\mbox{\small { \emph{first compute each $^l\!G$ with ports}}} & \qquad
\overline{c}_{11}X_1.\cdots.\overline{c}_{1n}X_{n}\par \para ^1\!G_{c_{11}\dots c_{1n}}^{b_1}\para\dots \\
\mbox{\small {  $c_{l1}\dots c_{ln}$ \emph{and} $b_l$, }} & \qquad
                                                                 \overline{c}_{k1}X_1.\cdots.\overline{c}_{kn}
                                                                 X_{n}\par \para ^k\!G_{c_{k1}\dots c_{kn}}^{b_k}\para\\
\mbox{\small { \emph{ then compute $F$ and output }}} & \qquad F_{b_1\dots b_k}^{b}) \\
\hline
\end{array}
\]

\item Recursion.  Suppose  $F$ is an $n$-ary recursive function, $G$ is an ($n{+}2$)-ary recursive function. Then the ($n{+}1$)-ary function $H$ defined in the following way is a recursive function: $H(\ve{x},0) = F(\ve{x}),\; H(\ve{x},n+1)  =  G(H(\ve{x},n),\ve{x},n) $. Then we have its $\Pi^{d}$ interpretation $Rec(F,G)_{a_1,\dots, a_n , a }^{b}$ as
\[\begin{array}{cl}\hline
&  \qquad Rec(F,G)_{a_1 \dots a_n, a}^{b} \DEF \\
& \qquad (c_1\dots c_n,d,e,f,g,h)a_1(X_1).\dots.a_n(X_n). a(X) \\
\mbox{\small { \emph{initialize}}} & \qquad ( !\overline{c}_1X_1.\dots.\overline{c}_nX_n ~|~ \overline{d}X ~|~ F_{c_1\dots c_n}^f ~|~ \\
\mbox{\small { \emph{increase from zero and try}}} & \qquad \overline{g}\enc{\underline{0}} ~|~ !Suc_e^g ~|~ \\
\mbox{\small { \emph{invariance} ($Y{+}Z=n_{a}$)}}& \qquad !(i,j,k)f(\stress{X}).g(\stress{Y}).d(\stress{Z}).(\stress{Z} \lrangle{i,j} ~|~\\
\mbox{\small { \emph{non-zero, call $G$}}}& \qquad i(\stress{Z'}).(G_{h,c_1\dots c_n,k}^f~|~ \overline{k}\stress{Y}.\overline{e}\stress{Y} ~|~ \overline{h}\stress{X}  ~|~\overline{d}\stress{Z'}) ~|~ \\
\mbox{\small { \emph{zero}}}& \qquad j.\overline{b} \stress{X}))\\\hline
\end{array}
\]

\item Minimization. Given  an $(n{+}1)$-ary recursive function $F$, the $n$-ary function $\mu y(F(\tilde{x},y)=0 )$ is also a recursive function.
    With the strategy of encoding recursion available, the minimization should be easy to understand.
\[ \begin{array}{cl}\hline
& \qquad {\mu(F)_{a_1\dots a_n}^{b}} \DEF  \\
& \qquad                                   (c_1\dots c_n, d,e,f,g,s)a_1(X_1).a_2(X_2).\dots.a_n(X_n) \\
\mbox{\small {\emph{initialize}}}&\qquad (\overline{c}_1X_1.\overline{c}_2X_2.\dots
                                   \overline{c}_nX_n.\overline{d}\enc{\underline{0}}
                                   ~|~!F_{c_1\dots, c_n,d}^f ~|~ \\
\mbox{\small {\emph{increase and try again}}} &\qquad !(s.Suc_e^g ~|~ g(X).\overline{d}X.\overline{e}X) ~|~ \overline{e}\enc{\underline{0}} ~|~ \\
\mbox{\small {\emph{test the value of $F$}}}&\qquad !(i,j)f(Y)(Y \lrangle{i,j}  ~|~ \\
\mbox{\small {\emph{non-zero}}}&\qquad i.\overline{s}.\overline{c}_1X_1.\overline{c}_2X_2.\dots \overline{c}_nX_n ~|~ \\
\mbox{\small {\emph{zero, output}}}&\qquad j.e(X).\overline{b}X ))) \\\hline
\end{array}
\]
\end{enumerate}

The following proposition establishes the relationship between the recursive functions and its interpretation. Intuitively, it says that they can always deliver the \emph{same} results to their environment.
\begin{proposition}\label{p:c1}
For a $k$-ary  recursive function $f(x_1,\ldots,x_k)$, let $F_{a_{1},\ldots,a_{k}}^{b}$ be its interpretation in $\Pi^{d}$ according to the above schema. Then on any input $(n_{1},\ldots, n_{k})$, if $f(n_{1},\ldots, n_{k})=m$ then

 $$(a_1,\ldots, a_k)(F_{a_{1},\ldots,a_{k}}^{b}~|~\overline \overline{{a_1}}(\enc{\underline{n_1}})~|~\ldots~|~\overline \overline{{a_k}}(\enc{\underline{n_k}}))~\strs{\AEHOPiddiv}~ \overline{b}(\enc{\underline{m}});$$

 and if $f(n_{1},\ldots, n_{k})$ is undefined then

  $$(a_1,\ldots, a_k)(F_{a_{1},\ldots,a_{k}}^{b}~|~\overline \overline{{a_1}}(\enc{\underline{n_1}})~|~\ldots~|~\overline \overline{{a_k}}(\enc{\underline{n_k}}))~\strs{\AEHOPiddiv} ~!\tau.$$
\end{proposition}

The proof of the proposition above is by induction on the construction of the recursive functions.
The proposition below states that the property of the above interpretation schema (note here we make a little abuse of the word \emph{encode} as we will show that  actually the above interpretation satisfies all but the back operational correspondence).
\begin{proposition}\label{p:c2}
There is a sound and complete encoding from $\mathbb{C}$ to $\Pi^d$.
\end{proposition}

 \strs{The soundness and completeness of the above encoding schema are examined with respect to $\equiv_{\mathbb{C}}$ and $\AEHOPiddiv$}.
Soundness follows easily from the fact that if two $\mathbb{C}$ processes are structural congruent then their encodings are bisimilar. For completeness, suppose for a contradiction that there exist some $P,Q$ such that $\encoding{P}{}{} \AEHOPid \encoding{Q}{}{}$ but $P\not\equiv_{\mathbb{C}} Q$.  As
the encoding scheme has ensured that the encoding does not introduce extra divergence,  from
$P\not\equiv_{\mathbb{C}} Q$ we know that there exists a process $E$ of $\mathbb{C}$-calculus
s.t. $P|E\Longrightarrow U$, where $U$ is an unobservable process ($0$ or $\Omega$), while $Q|E$
will not evolve to an unobservable process. Then it would follow that $\encoding{P|E}{}{}
\not\AEHOPid \encoding{Q|E}{}{}$, which implies $\encoding{P}{}{} \not\AEHOPid \encoding{Q}{}{}$
and contradicts the precondition. More discussion about $E$ and $U$ can be found in the proof of
Theorem \ref{th-C}. Finally, according to the encoding, every computation step taken by some
recursive function is realized by finite many internal steps, so the encoding is
divergence-reflecting.


It is straightforward to see that the encoding also satisfies all the static criteria: it is a structural encoding and does not introduce any new free names other than the (outmost) input/output channel names. The validity of the forth operational correspondence follows directly from the encoding scheme, which utilizes local names to realize the computation stepwise. The silent actions involved in the realization do not change the state of the processes (i.e. the intermediate processes are equivalent). However, generally the back operational correspondence would not be true. For example, for some $\mathbb{C}$ process $P$ with the input ability, one cannot predicate what might be transferred through the corresponding input channel of its encoding  $\enc{P}$. Any $\Pi^{d}$ process could be the object of the input, not just those which encode natural numbers. Such kind of inputs could render $P$ unable to match the subsequent steps of $\enc{P}$.

\begin{remark}
 As a matter of fact, the above encoding satisfies the so-called \emph{subbisimilarity} relation in \cite{Fu11a}, where one finds a comparative work about the computation completeness of $\pi$, and from which our encoding borrows some idea.
\end{remark}


\section{Expressiveness}\label{s:expressiveness}
This section formulates two encodings between $\pi$ and $\Pi^d$. We will explain the encoding strategy (with examples) and discuss their properties.

\subsection{Encoding $\pi$ in \HOPid}\label{s:expr:pi2Pid}
A central point here is how to mimic a $\pi$ name (say $u$), which may engage in both input and output. To do this, we consult the encoding idea from \cite{Tho93} (the relationship between name-parameterization and relabelling remains open however), 
and use the following gadget called \emph{pipe} that is a 3-ary abstraction.
\[\PIPEA{u} \DEF \lrangle{x_1,x_2,x_3}(x_1.u(Z).\overline{x_3}Z.0 \para x_2.x_3(Z).\overline{u}Z.0) \qquad x_1,x_2,x_3 \mbox{ fresh }
\] Intuitively, $x_1,x_2$ are to be instantiated by some local constants for receiving signals respectively indicating whether an input or output action is going to happen on the pipe. Parameter $x_3$ is to be instantiated by an auxiliary local name for ferrying the content of a prospective communication through the pipe. 
The encoding is illustrated in Figure \ref{pirep2Pidef}. 
\begin{figure}[thbp]
\noindent\rule{\textwidth}{.5pt}
\centering
\[
\begin{array}{lcl}
 \encoding{0}{1}{} &\DEF & 0 \\
 \encoding{u(x).P}{1}{} &\DEF & (i)(o)(c)(X_u\lrangle{i,o,c} \para \overline{i}.c(X_x).\encoding{P}{1}{}) \\
 \encoding{\overline{u}v.P}{1}{} &\DEF & (i)(o)(c)(X_u\lrangle{i,o,c} \para \overline{o}.\overline{c}X_v.\encoding{P}{1}{}) \\
 \encoding{P\para Q}{1}{} &\DEF & \encoding{P}{1}{}\para \encoding{Q}{1}{}\\
 \encoding{(d)P}{1}{} &\DEF & (d)(\encoding{P}{1}{}\hosub{\PIPEA{d}}{X_d}) \\
 \encoding{!P}{1}{} & \DEF& !\encoding{P}{1}{} \;(!P \mbox{ is guarded replication})
\end{array}
\]
\[
\begin{array}{lcl}
\encoding{P}{}{} &\DEF & \encoding{P}{1}{}\hosub{\PIPEA{a_1}}{X_{a_1}}\cdots\hosub{\PIPEA{a_n}}{X_{a_n}}
\end{array}
\]
\noindent\rule{\textwidth}{.5pt}
\caption{Encoding from $\pi$ to \HOPid}\label{pirep2Pidef}
\end{figure}
Every $\pi$ name is corresponded to a process variable. Specifically, we assume a one-to-one mapping from $n(P){=}\{u_1,...,u_k\}$ to $\{X_{u_1},...,X_{u_k}\}{=}v(\encoding{P}{1}{})$. We thus have the following corresponding in the encoding. 
\begin{itemize}
\item Every free name constant (e.g. $a$) is mapped to a free process variable (e.g. $X_a$) in $\encoding{}{1}{}$ that is then instantiated by the pipe (e.g. $\PIPEA{a}$) corresponding to this free name constant in $\encoding{}{}{}$.
\item Every bound name constant (e.g. $d$ in $(d)P$) is mapped to a free process variable (e.g. $X_d$) that is instantiated by the pipe (e.g. $\PIPEA{d}$) corresponding to this bound name constant in $\encoding{}{1}{}$, where the name $d$ in the pipe (e.g. $\PIPEA{d}$) is still bound.
Bound/Local $\pi$ name constants are allotted pipes in $\encoding{}{1}{}$ rather than in $\encoding{}{}{}$ because they are essentially distinct from free name constants.
\item Every (input-prefix) bound name variable (e.g. $x$) is mapped to a process variable (e.g. $X_x$) in $\encoding{}{1}{}$, where the process variable (e.g. $X_x$) is bound and to be instantiated by an inputted process (i.e. a pipe).
Note an input-bounded name variable in $P$ is not replaced immediately since it must be determined through communication.
\end{itemize}
The part needing explanation is those concerning the prefixes. In principle, prefixes are interpreted in the following way: one first decides how to use the pipe (i.e. input or output); then through several internal communication on the auxiliary names ($i,o,c$), the first-order communication in $\pi$ is encoded by a communication in \HOPid.
For instance, in $\encoding{u(x).P}{1}{}$, $u$ maps to $X_u$ that is to be substituted by $\PIPEA{u}$, is notified on $i$ that an input action is engaged, then the real input action on $u$ occurs in $\PIPEA{u}$, and the received process (another pipe) is relayed through $c$ to the rest of the process undergoing encoding.
We present an example in \cite{XYL15} (for the sake of space).

\paragraph{Properties of the encoding}
One can see that the static conditions are naturally met.
Then an observation is the following lemma (Lemma \ref{eg:encodinginho}) whose proof is a structure induction.
\begin{lemma}\label{eg:encodinginho}
The following action correspondence from the encoding holds:
(1) If $P\st{a(b)} P'$, then \\ $\encoding{P}{}{}\st{\tau}\st{a(\PIPEA{b})}\st{\tau} \SAEHOPid \encoding{P'}{}{}$;
(2) If $P\st{\overline{a}b} P'$, then $\encoding{P}{}{}\st{\tau}\st{\tau}\st{\overline{a}[\PIPEA{b}]} \SAEHOPid \encoding{P'}{}{}$;
(3) If $P\st{\overline{a}(b)} P'$, then \\ $\encoding{P}{}{}\st{\tau}\st{\tau}\st{(b)\overline{a}[\PIPEA{b}]} \SAEHOPid \encoding{P'}{}{}$;
(4) If $P\st{\pmb{\tau}} P'$, then
$\encoding{P}{}{}\st{\tau}\st{\tau}\st{\tau}\st{\pmb{\tau}}\st{\tau} \SAEHOPid \encoding{P'}{}{}$.
\end{lemma}
Notice that in Lemma \ref{eg:encodinginho}(4), the two $\pmb{\tau}$'s exactly correspond with each other, which means that in the five consecutive $\tau$ actions of $\encoding{P}{}{}$, the fourth $\tau$ is the one that actually comes from process $P$, and the others are auxiliary ones brought about by the encoding.
Based on this observation and a (transition) induction, we have the operational correspondence (Lemma \ref{fo2ho-ac} and Lemma \ref{fo2ho-cac}).
\begin{lemma}\label{fo2ho-ac}
Suppose $P,Q$ are $\pi$ processes. Then the following properties hold:\\
(1) If $P\wt{a(b)} P'$, then $\encoding{P}{}{} \wt{a(\PIPEA{b})} \SAEHOPid \encoding{P'}{}{}$;\\
(2) If $P\wt{\overline{a}b} P'$, then $\encoding{P}{}{} \wt{\overline{a}[\PIPEA{b}]} \SAEHOPid \encoding{P'}{}{}$;\\
(3) If $P\wt{\overline{a}(b)} P'$, then $\encoding{P}{}{} \wt{(b)\overline{a}[\PIPEA{b}]} \SAEHOPid \encoding{P'}{}{}$;\\
(4) If $P\wt{\tau} P'$, then $\encoding{P}{}{} \wt{\tau}  \SAEHOPid \encoding{P'}{}{}$.
\end{lemma}
\begin{lemma}\label{fo2ho-cac}
Suppose $P,Q$ are $\pi$ processes. The following properties hold:\\
(1) If $\encoding{P}{}{} \wt{a(\PIPEA{b})} T$, then $P\wt{a(b)} P'$ for some $P'$ and $T \AEHOPid \encoding{P'}{}{}$; \\
(2) If $\encoding{P}{}{} \wt{\overline{a}[\PIPEA{b}]} T$, then $P\wt{\overline{a}b} P'$ for some $P'$ and $T \AEHOPid \encoding{P'}{}{}$; \\
(3) If $\encoding{P}{}{} \wt{(b)\overline{a}[\PIPEA{b}]} T$, then $P\wt{\overline{a}(b)} P'$ for some $P'$ and $T \AEHOPid \encoding{P'}{}{}$; \\
(4) If $\encoding{P}{}{} \wt{\tau} T$, then: if the weak transition contains solely silent actions on the auxiliary names (i.e. $i',o',c'$) it holds $T \AEHOPid \encoding{P}{}{}$; otherwise $P\wt{\tau} P'$ for some $P'$ and $T \AEHOPid \encoding{P'}{}{}$.
\end{lemma}

\paragraph{Completeness} The encoding is complete with respect to $\AEFOPi$ and $\AEHOPid$.
\begin{lemma}\label{pi2Pi-full-ab-complete}
Let $P,Q$ be $\pi$ processes. $\encoding{P}{}{}\AEHOPid \encoding{Q}{}{} \mbox{ implies } P \AEFOPi Q$.
\end{lemma}
To prove Lemma \ref{pi2Pi-full-ab-complete}, we firstly use the `total' property of the encoding. i.e., every `context' $C$ of $\pi$, going through the encoding, is corresponded by a context $D\equiv \encoding{C}{}{}$ in \HOPid. Secondly, we take advantage of the characterization of weak bisimilarity in $\pi$ (i.e. local bisimilarity).
We detail the proof in \cite{XYL15}.

\paragraph{Soundness} One would expect the encoding is sound with respect to $\AEFOPi$ and $\AEHOPid$. Unfortunately, this is still unknown, but we do have somewhat an approximation.
\begin{lemma}\label{pi2Pi-full-ab-sound}
Let $P,Q$ be $\pi$ processes. $P \AEFOPi Q \mbox{ implies } \encoding{P}{}{}\approx_{p} \encoding{Q}{}{}$.
\end{lemma}
The equivalence $\approx_{p}$ is defined as below.
\begin{definition}
A symmetric relation $\mathcal{R}$ on \HOPid~processes is a pipe-bisimulation, if whenever $P\mathcal{R} Q$, the following properties hold.
\begin{itemize}
\item If $P \st{a(\PIPEA{f})} P'$, then $Q \wt{a(\PIPEA{f})} Q'$ for some $Q'$ and $P'\mathcal{R} Q'$;
\item If $P \st{(\ve{c})\overline{a}[\PIPEA{f}]} P'$ in which $\ve{c}$ is $f$ or empty, then $Q \wt{(\ve{c})\overline{a}[\PIPEA{f}]} Q'$ for some $Q'$, and for every process $E[X]$ s.t. $\ve{c}\cap fn(E)=\emptyset$ it holds that
$(\ve{c})(E[\PIPEA{f}]\para P') \; \mathcal{R}\;  (\ve{c})(E[\PIPEA{f}]\para Q')$.
\item If $P \st{\tau} P'$, then $Q \wt{} Q'$ for some $Q'$ and $P'\mathcal{R} Q'$;
\end{itemize}
Pipe-bisimilarity $\approx_{p}$ is the largest pipe-bisimulation.
\end{definition}
To prove Lemma \ref{pi2Pi-full-ab-sound}, a crucial point is on the matching of higher-order input. That is, we need to prove the following (notice a dotted line means the two processes connected vertically by it are related by the relation marked beside the line, and the squiggly arrow reads `implies').
\[
\xymatrix{
 \encoding{P}{}{} \ar@{.}[d]_{\mathcal{R}}\ar@{=>}[rr]^{\hspace*{-.9cm} a(\PIPEA{d})}  & & S_1\hosub{\PIPEA{d}}{X}\ar@{.}[d]_{\Delta}\ar@{~>}[r]^{} & P\ar@{=>}[r]^{a(d)}\ar@{.}[d]^{\AEFOPi} & P'\ar@{.}[d]^{\AEFOPi}\\ 
\encoding{Q}{}{}\ar@{=>}[rr]^{\hspace*{-.9cm} a(\PIPEA{d})} & & T_1\hosub{\PIPEA{d}}{X} & Q\ar@{=>}[r]^{a(d)}\ar@{~>}[l]^{} & Q' 
}
\]
where $\mathcal{R} \DEF \{ (\encoding{P}{}{},\encoding{Q}{}{}) \;|\; P \AEFOPi Q \}$.
Then one can derive $\Delta$ which indicates the relating of $S_1\hosub{\PIPEA{d}}{X}$ and $T_1\hosub{\PIPEA{d}}{X}$ by $\mathcal{R}$.
That said, a desired result would be that $\approx_{p} \subseteq \AEHOPid$ (that $\AEHOPid \subseteq \approx_{p}$ is obvious). Yet this remains an open issue.

\paragraph{Divergence-reflecting} The encoding is not divergence-reflecting because in the encoding of the replication a prefix can wait for an infinite number of $\tau$ actions occurring on auxiliary names ($i,o,c$), thus introducing divergence. However this can be fixed by adjusting the definition of replication as below (we define the case of input-guarded replication, and output-guarded replication is similar). This adjustment simply postpones the production of another copy of the replication until the visible action fires, and thus would not change other properties.
\[
\begin{array}{rcl}
\enc{!u(x).P} &\DEF& (d)(Q_d \para \overline{d}Q_d) \mbox{ where } \\
Q_d &\DEF& d(Z).((i)(o)(c)(X_u\lrangle{i,o,c} \para \overline{i}.c(X_x).(Z\para \enc{P}) \para \overline{d}Z)
\end{array}
\]

\subsection{Encoding \HOPid~ in $\pi$}\label{s:expr:Pid2pi}
This section presents an encoding from $\Pi^d$ to $\pi$.
The encoding exploits the idea of \emph{triggers} \cite{San92a}\cite{San92}\cite{SW01a}.
The difference from the encodings in these work is three-fold: First, $\Pi^d$ is a purely higher-order calculus (i.e. no name-passing) and has solely parameterization on names (not involving parameterization on processes); second, $\pi$ has no parameterization itself; third, $\Pi^d$ has general name-parameterization (i.e. arbitrary arity). 
\begin{figure}[tbhp]
\noindent\rule{\textwidth}{.5pt}
\centering
\[
\begin{array}{lcl}
 \encoding{0}{}{} &\DEF&  0 \\
 \encoding{u(X).P}{}{} &\DEF& u(x).\encoding{P}{}{}\\
 \encoding{\overline{u}Q.P}{}{} &\DEF& (f)(\overline{u}f.\encoding{P}{}{}\para T), \\
 && T\DEF
 \left\{\begin{array}{l}
  !f(z).z(x_1).\cdots. z(x_n).\encoding{Q'}{}{} \quad \mbox{ if } Q\equiv \lrangle{x_1,...,x_n}Q' \\
  !f.\encoding{Q'}{}{} \quad \mbox{ otherwise } (n=0)
 \end{array}\right.\\
 && T \mbox{ is sometimes simply abbreviated as } !\ve{f(x)}.\encoding{Q'}{}{}  \\
 \encoding{P\para Q}{}{} &\DEF & \encoding{P}{}{}\para \encoding{Q}{}{}\\
 \encoding{(c)P}{}{} &\DEF & (c)\encoding{P}{}{} \\
 \encoding{\lrangle{x_1,...,x_n}P}{}{} &\DEF & 0 \\ 
 \encoding{X\lrangle{d_1,...,d_n}}{}{} &\DEF &
 \left\{\begin{array}{l}
 \overline{x}(g).\overline{g}d_1.\cdots.\overline{g}d_n \quad \mbox{ if } n\neq 0 \\
 \overline{x}\quad \mbox{ otherwise }
 \end{array}\right.
\end{array}
\]
\noindent\rule{\textwidth}{.5pt}
\caption{Encoding from \HOPid~to $\pi$}\label{Pidef2pi}
\end{figure}

We provide the encoding in Figure \ref{Pidef2pi}. 
In the encoding, we assume every name variable in a \HOPid~process is mapped to the same variable in $\pi$, for example, $x$ in $\lrangle{x}P$ to $x$; and every process variable in a \HOPid~process is mapped to a name variable (in lowercase), for example, $X$ to $x$. We assume $\alpha$-conversion is applied if needed.
The central part of the encoding strategy resides on how to translate the scenario concerning the communication of a parameterized process, i.e. the transmission of a process of the form $\lrangle{\ve{x}}P$ that is going to instantiate a variable, say $X$, in the receiving environment, which has an input-guarded (sub)process like $X\lrangle{\ve{d}}$ that then undergoes the application on the parameterized names.

In the translation of output, the internal channel $f$ (the trigger name) is the access point 
to the (parameterized) process to be transmitted, whose instantiation of the parameters is realized through  $f$ (and $g$), whenever an application happens.
Notice that the process transmitted, of the form $\lrangle{\ve{x}}P$ in general, is turned into an open process $\encoding{P}{}{}$, and thus to receive the constant names through $f$ (and $g$) to instantiate the name variables in $\encoding{P}{}{}$.
During the reception of the constant names, the private channel established on local name $g$ is used to ensure no intervention will occur when encoding a process like $a(X).(X\lrangle{d_1}\para X\lrangle{d_2})$, and this is akin to translating polyadic communication in a sense.
There is a special case that $Q$ is not a parameterized process ($\size{\ve{x}}=0$).

Moreover, merely $X\lrangle{\ve{d}}$ is defined rather than the general form $P\lrangle{\ve{d}}$, because in the latter the application can be immediately fired if the process $P$ is correctly defined, resulting in a process not parameterized outmost. The definition of the encoding of $X\lrangle{\ve{d}}$ is to successively convey, through $g$, the constant names for instantiating the name variables in a prospective communicated process at the access point $x$, which will be instantiated by a trigger name (e.g. $f$). A special case is $n=0$, when $X$ is simply encoded as $\overline{x}$.
The rest part of the encoding is defined in a homomorphic way.
To help understand the encoding, we provide an example in \cite{XYL15}.

\paragraph{Properties of the encoding}
A straightforward observation of the encoding is that it is compositional and name-preserving.
Meanwhile the encoding is divergence-reflecting because it does not introduce divergence.
Lemma \ref{Pid2piOpCor} and Lemma \ref{Pid2piOpCor-con} clarify the operational correspondence of the encoding. Their proofs are by induction on the transition of $P$ (or $\encoding{P}{}{}$). 
\begin{lemma}\label{Pid2piOpCor}
 Let $P$ be a \HOPid~process. 
Notice that $\ve{f(x)}$ is  $f(z).z(x_1).\cdots. z(x_n)$ if $A\equiv \lrangle{x_1,x_2,...,x_n}A'$, or $f$ if $n=0$ (i.e. A is not parameterized).
\begin{enumerate}
\item[(1)] If $P\wt{(\ve{c})\overline{a}A} P'$, then $\encoding{P}{}{} \wt{\overline{a}(f)} \SAEFOPi 
 (\ve{c})(\encoding{P'}{}{}\para !\ve{f(x)}.\encoding{A'}{}{})$. 
\item[(2)] If $P\wt{a(A)} P'$, and thus there exists some $E[X]$ s.t. $P'\equiv E[A]$.\\
Then $\encoding{P}{}{} \wt{a(f)} \SAEFOPi \encoding{E}{}{}\fosub{f}{x}$  
and $(f)(\encoding{E}{}{}\fosub{f}{x} \para !\ve{f(x)}.\encoding{A'}{}{}) \AEFOPi \encoding{P'}{}{}$.
\item[(3)] If $P\wt{\tau} P'$, then there exists $T$ s.t.
 $ \encoding{P}{}{} \wt{\tau} T \AEFOPi \encoding{P'}{}{} $
\end{enumerate}
\end{lemma}
\begin{lemma}\label{Pid2piOpCor-con}
 Let $P$ be a \HOPid~process. 
Notice that $\ve{f(x)}$ is  $f(z).z(x_1).\cdots. z(x_n)$ if $A\equiv \lrangle{x_1,x_2,...,x_n}A'$, or $f$ if $n=0$ (i.e. A is not parameterized).
\begin{enumerate}
\item[(1)] If $\encoding{P}{}{} \wt{\overline{a}(f)} T$, then $P\wt{(\ve{c})\overline{a}A} P'$ for some $A,P',\ve{c}$, and $T\AEFOPi \;(\ve{c})(\encoding{P'}{}{}\para !\ve{f(x)}.\encoding{A'}{}{})$.
\item[(2)] If $\encoding{P}{}{} \wt{a(f)} T$,
then $P\wt{a(A)} E[A]$ for some $E[X]$ s.t. $T\AEFOPi \encoding{E}{}{}\fosub{f}{x}$,
and \\
$(f)(T \para !\ve{f(x)}.\encoding{A'}{}{}) \AEFOPi \encoding{E[A]}{}{}$.
\item[(3)] If $\encoding{P}{}{}\wt{\tau} T$, then $P\wt{\tau} P'$ for some $P'$, and $\encoding{P'}{}{} \AEFOPi T$.
\end{enumerate}
\end{lemma}

\paragraph{Completeness and soundness} The encoding is complete and sound with respect to $\AEHOPid$ and $\AEFOPi$.
\begin{lemma}\label{prop-full-abstr1-abstr2}
Let $P,Q$ be \HOPid~processes. $P \AEHOPid Q  \mbox{ iff } \encoding{P}{}{} \AEFOPi \encoding{Q}{}{}$.
\end{lemma}
The proof of completeness can be done by showing the relation $\{(P, Q) \;|\;  \encoding{P}{}{} \AEFOPi \encoding{Q}{}{}\} \cup \AEHOPid$ to be a context bisimulation up-to $\AEHOPid$, using Lemma \ref{Pid2piOpCor} and Lemma \ref{Pid2piOpCor-con} to move back and forth in action correspondence.
The proof of soundness is a bit more complicated and can be conducted by showing the following relation $\{(\encoding{P}{}{}, \encoding{Q}{}{}) \;|\; P \AEHOPid Q\}\,\cup\AEFOPi$ to be a weak bisimulation up-to injective substitution and $\AEFOPi$ \cite{SW01a}, with the help of Lemma \ref{Pid2piOpCor} and Lemma \ref{Pid2piOpCor-con} in pinpointing the action before and after the encoding. The up-to technique is a method of simplifying the design of the relation toward proving bisimulation equivalence (see \cite{SW01a}\cite{San98}).
Base upon the discussion above, we have the following proposition.
\begin{proposition}
There is an encoding from \HOPid~to $\pi$.
\end{proposition}


\section{Conclusion}\label{s:conclusion}
In this paper, we have studied from two computation angles the name parameterization in a higher-order setting.  First is about the computation completeness. We show that there is a sound and complete embedding of an elementary model $\mathbb{C}$ with built-in recursive functions into the higher-order pi-calculus with name parameterization. This method actually can be extended to other concurrency formalism. Second is about relative expressiveness. We provide the mutual encodings between $\pi$ and higher-order pi-calculus with name parameterization. We discuss the properties with respect to well-known criteria for encodings. They offer insight into the position of these calculi in the expressiveness hierarchy as well as some programming skills. Some further work include, among others: examining further to refine the properties concerning the encoding from $\pi$ to $\Pi^d$; studying the computation completeness of $\Pi^D$ (i.e. higher-order pi-calculus with parameterization on processes themselves) and comparing it further with $\Pi^d$. 


\section*{Acknowledgement}
This work has been supported by project ANR 12IS02001 PACE and NSF of China (61261130589, 61173048, 61202023, 61472239). The authors are also grateful to the comments and suggestions from the anonymous referees.


\bibliographystyle{eptcs}
\bibliography{process}


\end{document}